\documentclass[preprint,showpacs,preprintnumbers,amsmath,amssymb,natbib]{revtex4}

\usepackage{graphicx}
\usepackage{epsfig}		
\usepackage{dcolumn}
\usepackage{bm}
\def\0{\mbox{\tiny $0$}}
\def\1{\mbox{\tiny $1$}}
\def\2{\mbox{\tiny $2$}}
\def\3{\mbox{\tiny $3$}}
\def\4{\mbox{\tiny $4$}}
\def\5{\mbox{\tiny $5$}}
\def\6{\mbox{\tiny $6$}}
\def\7{\mbox{\tiny $7$}}
\def\8{\mbox{\tiny $8$}}
\def\9{\mbox{\tiny $9$}}

\def\kk{\mbox{\small $k$}}

\def\f14{\mbox{\tiny $\frac{1}{4}$}}

\def\F{\mbox{\tiny $F$}}

\def\ii{\mbox{\tiny $i$}}
\def\jj{\mbox{\tiny $j$}}
\def\kk{\mbox{\tiny $k$}}

\def\s{\mbox{\tiny $s$}}

\def\mi{\mbox{\tiny $-$}}

\def\bb#1{\mbox{\footnotesize $(#1)$}}

\begin{document}

\title{Degenerate Fermi gas perturbations at standard background cosmology}

\author{A. E. Bernardini}
\email{alexeb@ufscar.br, alexeb@ifi.unicamp.br}
\affiliation{Departamento de F\'{\i}sica, Universidade Federal de S\~ao Carlos, PO Box 676, 13565-905, S\~ao Carlos, SP, Brasil}
\author{E. L. D. Perico}
\email{elduarte@ifi.unicamp.br}
\affiliation{Instituto de F\'{\i}sica Gleb Wataghin, Universidade Estadual de Campinas, PO Box 6165, 13083-970,
Campinas, SP, Brasil}

\date{\today}

\begin{abstract}
The hypothesis of a tiny fraction of the cosmic inventory evolving cosmologically as a degenerate Fermi gas test fluid at some dominant cosmological background is investigated.
Our analytical results allow for performing preliminary computations to the evolution of perturbations for relativistic and non-relativistic test fluids.
The density fluctuation, $\delta$, the fluid velocity divergence, $\theta$, and an explicit expression for the dynamics of the shear stress, $\sigma$, are obtained for a degenerate Fermi gas in the background regime of radiation.
Extensions to the dominance of matter and to the $\Lambda$CDM cosmological background are also investigated and lessons concerning the formation of large structures of degenerate Fermi gas are depicted.
\end{abstract}

\pacs{98.80.-k, 03.75.Ss, 98.70.Sa}
\keywords{}
\date{\today}
\maketitle

\section{Introduction}

The complete understanding of the cosmic inventory is one of the most challenging issues of theoretical physics.
The efforts to relate theory with the observational data and to achieve an enlarged overview of the dark sector do necessarily require an interplay between general relativity, astrophysics and particle physics \cite{Teo2,Teo3,Teo4,Teo1}.
It may provide guidelines to the resolution of some fundamental questions that indeed influence the development of particle physics beyond its standard model (SM).

Our proposal concerns the investigation of matter components evolving cosmologically as a degenerate Fermi gas (DFG).
We apply the linear cosmological perturbation theory to obtain an analytical description of the growth rate of perturbations for a DFG test fluid.
It is suggested that particles behaving as a DFG are related to a tiny fraction of dark matter (in fact, less than $1\%$) that could be made of massive neutrinos \cite{Ber10} in adiabatic equilibrium with the radiation background \cite{Bil97}.
In fact, just under very particular circumstances it provides attractive conditions for a model of dark matter components \cite{Ber09,Ber09B}.
In general, neutrinos in the cosmic background are treated either in the limit of radiation regime or in the limit of cold matter. Treating massive neutrinos as a DFG allows for quantifying a smooth transition between ultra-relativistic (UR) and nonrelativistic (NR) thermal regimes.
It becomes a convenient tool used for describing the role of such particles in the cosmic inventory, from radiation-dominated (RD) to matter-dominated (MD) eras.

In the context of the standard cosmology, theoretical speculations about the existence of a degenerate Fermi gas (DFG) of neutrinos produced in the early universe were carried out in the past \cite{Zel81}.
Depending on the value of the mass, the density of neutrinos is enhanced when they cluster into gravitational potential
wells, although that overdensity is limited to small factors \cite{Past01}.
Otherwise, the role of neutrinos as hot dark matter (HDM) particles with a large velocity dispersion has been perpetuated in the literature \cite{Past02,Past03,Past04}.
The approach here introduced allows for parameterizing the transition between HDM and cold dark matter (CDM) through the analytical expressions for energy density and pressure of a DFG.

The formalism for a DFG describes a canonical ensemble that, when cooled below a critical temperature, presents a condensed phase consisting of {\em quasidegenerate fermion lump} \cite{Bil99,Pao01,Mun02}.
Self-gravitating fermion systems, for instance, could be conceived through the introduction of a distribution function modified with an energy cutoff in phase space, as it happens for a DFG \cite{Kremer2002,Stella1983,Chava2002}.
It has also been showed that for a self-gravitating Fermi gas at finite temperature, below a critical temperature, it undergoes a first order phase transition leading to a compact object with a degenerate core.
It is expected that at some stage of the evolution of the Universe, primordial density fluctuations have become gravitationally unstable forming dense clumps of dark matter that have survived in the form of galactic halos \cite{Boy10}.

The possibilities of forming large structures have stimulated us to investigate the behavior of a DFG test fluid in the cosmological background of radiation, matter and $\Lambda$CDM.
Due to phenomenological reasonable arguments, one should expect a small contribution from such test fluid to the cosmic dynamics.
In addition, the procedure described in this work is complementary to previous studies of massive neutrinos in cosmology, where the study of NR and UR fluids (i. e. massive and massless neutrinos) \cite{Ma94,Pas06,Dolgov02} had to be performed separately from the beginning of analysis.

To study the cosmological evolution of a DFG as a test fluid in some dominant cosmological background, our manuscript was organized as follows.
In section II we report about the usual analytical approach for the Einstein equations for perturbations.
Part of our results are depicted from section II.
We show how to treat analytically the perturbations for relativistic and non-relativistic fluids based only on three realistic approximations.
In section III we adapt such results for obtaining the evolution equations for the density and fluid velocity perturbations of a DFG.
In section IV we obtain the preliminary results for density fluctuation, $\delta$, and fluid velocity divergence, $\theta$, for a DFG test fluid in the RD background universe.
Extensions to MD and $\Lambda$-dominated ($\Lambda$D) background scenarios are considered.
It consistently follows analytical continuity conditions.
We draw our conclusions in section V.

\section{Analytical approach for perturbations}

For a homogeneous Friedmann-Robertson-Walker (FRW) flat universe with energy density, $\bar{\rho}\bb{\tau}$, and pressure, $\bar{P}\bb{\tau}$, one has the following evolution equations for the scale factor, $a\bb{\tau}$,
\begin{eqnarray}
         \left( {\dot{a}\over a} \right)^{\2} &=& {8\pi\over3}G a^{\2} \bar{\rho} \,,\\
{d\over d\tau} \left( {\dot{a}\over a} \right)&=& -{4\pi\over 3}G a^{\2} (\bar{\rho}+3\bar{P}) \,
\label{friedmann}
\end{eqnarray}
where the dots denote derivatives with respect to the conformal time, $\tau$.
It follows that the expansion factor scales as $a\propto\tau$ in the RD era and as $a\propto\tau^{\2}$ in
the MD era.

We find that it is more convenient to solve the the first-order perturbed Einstein equations in the conformal Newtonian gauge, so that one has
\begin{mathletters}
\begin{eqnarray}
    k^{\2}\phi + 3{\dot{a}\over a} \left( \dot{\phi} + {\dot{a}\over a}\psi
	\right) &=& 4\pi G a^{\2} \delta T^{\0}_{\0} \,,
	\label{ein-cona}\\
    k^{\2} \left( \dot{\phi} + {\dot{a}\over a}\psi \right)
	 &=& 4\pi G a^{\2} (\bar{\rho}+\bar{P}) \theta
	 \,,\label{ein-conb}\\
    \ddot{\phi} + {\dot{a}\over a} (\dot{\psi}+2\dot{\phi})
	+\left(2{\ddot{a} \over a} - {\dot{a}^{\2} \over a^{\2}}\right)\psi
	+ {k^{\2} \over 3} (\phi-\psi)
	&=& {4\pi\over 3} G a^{\2} \delta T^{\ii}_{\ii}
	\,,\label{ein-conc}\\
    k^{\2}(\phi-\psi) &=& 12\pi G a^{\2} (\bar{\rho}+\bar{P})\sigma
	\,,\label{ein-cond}
\end{eqnarray}
\label{ein-con}
\end{mathletters}
where the variables $\theta$ and $\sigma$ are defined as
\begin{equation}
\label{theta}
 	(\bar{\rho}+\bar{P})\theta \equiv i k^{\jj} \delta T^{\0}_{\jj}\,,
	\qquad	(\bar{\rho}+\bar{P})\sigma \equiv -(\hat{k}_{\ii}\hat{k}^{\jj}
	- {1\over 3} \delta_{i}^{j})\Sigma^{\ii}_{\jj}\,,
\end{equation}
and $\Sigma^{\ii}_{\jj} \equiv T^{\ii}_{\jj}-\delta^{\ii}_{\jj} T^{\kk}_{\kk}/3$ denotes the traceless component of $T^{\ii}_{\jj}$.


For a perfect fluid of energy density $\rho$ and pressure $P$ at a comoving frame, the energy-momentum tensor has the form
\begin{equation}
	T^\mu_{\nu} = P g^\mu_{\nu} + (\rho + P) U^\mu U_\nu \,,
\end{equation}
where $U^\mu = dx^\mu /\sqrt{-ds^{\2}} $ is the four-velocity of the fluid.
For a fluid moving with a small coordinate velocity $v^{\ii} \equiv dx^{\ii}/d\tau$, $v^{\ii}$ can be treated as a perturbation like $\delta\rho=\rho-\bar{\rho}$, $\delta P=P-\bar{P}$, $\phi$ and $\psi$.
Then to linear order in the perturbations the energy-momentum tensor is given by
\begin{eqnarray}
	T^{\0}_{\0} &=& -(\bar{\rho} + \delta\rho) \,,\nonumber\\
	T^{\0}_{\ii} &=& (\bar{\rho}+\bar{P}) v_{\ii}  = -T^{\ii}_{\0}\,,\nonumber\\
	T^{\ii}_{\jj} &=& (\bar{P} + \delta P) \delta^{\ii}_{\jj}
		+ \Sigma^{\ii}_{\jj} \,,\qquad
		\Sigma^{\ii}_{\ii}=0 \,,
\end{eqnarray}
where we have allowed an anisotropic shear perturbation
$\Sigma^{\ii}_{\jj}$ in $T^{\ii}_{\jj}$.
At this point, it is pertinent to observe that, since the photons are tightly coupled to the baryons before recombination, the dominant contribution to this shear stress comes from the cosmological neutrinos.
One notices that, for a fluid, $\theta$ defined in equation (\ref{theta}) is simply the divergence of the fluid velocity: $\theta = i\, k^{\jj} v_{\jj}$.

Through the Bianchi identities, the Einstein equation implies the conservation of the total energy-momentum tensor.
Actually, the energy-momentum tensor of each uncoupled fluid is conserved, and obeys the continuity equation in $k$-space,
\begin{eqnarray}
\label{delta2}
	\dot{\delta} &=& - (1+w) \left(\theta-3{\dot{\phi}}\right),
\end{eqnarray}
and the Euler equation,
\begin{eqnarray}
\label{theta2}	\dot{\theta} &=& - {\dot{a}\over a} (1-3w)\theta - {\dot{w}\over
	     1+w}\theta + {w \over 1+w}\,k^{\2}\delta
	     - k^{\2} \sigma + k^{\2} \psi,
\end{eqnarray}
where we have assumed that, for isentropic primordial perturbations, $\delta P/\delta\rho = c^{\2}_s = P/\rho = w$, where $c_s$ is the adiabatic sound speed in the fluid.
In the non-relativistic fluid description for CDM and the baryon components, one sets $w \approx 0$.
The photon and the neutrino components should be more accurately described by their full distribution functions in phase space.
Herewith we shall develop the procedure for treating perturbations valid for relativistic and non-relativistic test fluids in a DFG regime.

To treat a DFG as a test fluid in some generical cosmological background, we have to follow a sequence of three approximations.
Let us consider that the phase space distribution of the particles gives the number of particles in a differential volume $dx^{\1} dx^{\2} dx^{\3} dp_{\1} dp_{\2} dp_{\3}$ in phase space,
\begin{equation}
f\bb{p\bb{E, n_{\jj}},  x^{\ii}, \tau}\,dx^{\1} dx^{\2} dx^{\3} dp_{\1} dp_{\2} dp_{\3}/(2\pi)^3 = dN.
\end{equation}
where $f$ is a Lorentz scalar that is invariant under canonical transformations.
Assuming that linear perturbations on the temperature is parameterized by
\begin{equation}
T\,(1 + \Delta\bb{x^{\ii}, n_{\jj}, \tau}),
\end{equation}
with $T\equiv T\bb{\tau} \equiv T\bb{a}$, the zeroth-order element of the phase space distribution, $f\bb{p\bb{E, n_{\jj}},  x^{\ii}, \tau}$, is the Fermi-Dirac distribution given by
\begin{equation}
f\bb{E, \tau} = g \, \left(\exp{\left[\frac{E - \mu}{T}\right]} + 1\right)^{\mi\1} = g \, \left(\exp{\left[\frac{E}{T} - \frac{\mu_{\0}}{T{\0}}\right]} + 1\right)^{\mi\1},
\label{dist}
\end{equation}
where the factor $g$ is the number of spin degrees of freedom, the Planck and the Boltzmann constants, $\hbar$ and $k$, were set equal to unity, and $T_{\0} = \mu_{\0} (T/\mu)$ denotes the temperature of the particles today.
In addition, we have introduced into the Eq.~(\ref{dist}) our {\em first approximation}.
The thermodynamics of our test fluid is constrained by the cosmological behavior for the chemical potential, $\mu\sim \mu\bb{a}$.
In the literature, the contribution due to the chemical potential $\mu$ to the distribution function is often discarded by setting $\mu = 0$.
Since we are interested in depicting the cosmological behavior of a DFG, we cannot perform the same approximation, i. e. we cannot set $\mu_{\0} = 0$.
The simplest way to manipulate it in the Boltzmann equation by keeping valid the condition
\begin{equation}
\left.\frac{\mbox{d} f}{\mbox{d} \tau}\right|_{\Delta \equiv 0} = \left.\frac{\mbox{d} f}{\mbox{d} a}\right|_{\Delta \equiv 0} = 0
\label{cond}
\end{equation}
is by assuming that $\mu/T = \mu_{\0}/T_{\0}$ does not depend on $a$, i. e. that the chemical potential follows the same temperature fluctuations on time.
Consequently it does not contribute to first-order corrections of the Boltzmann equation and the following constraint between $f$ and $T$ is maintained\footnote{In the same way, one could assume that $\mu$ has magnitude of first-order corrections to T. But it constrains $\mu$ to very small values, which could not be our hypothesis.},
\begin{equation}
\frac{\partial f}{\partial E} = \frac{E}{p} \frac{\partial f}{\partial p}
 = - \frac{1}{E} \left(T \frac{\partial f}{\partial T}\right).
\label{dist2}
\end{equation}
The Fermi momentum of a DFG will thus be written as $p_{F} = q_{F}/a$ and its comoving representation as $q_{F} = \sqrt{\mu_{\0}^{\2} - m^{\2}a^{\2}}$, i. e. the ratio $q_{F}/m$ diminishes as the Universe expands.
If the chemical potential at present, $\mu_{\0}$, is larger than the particle mass, $m$, of the test fluid, it is possible to depict analytically the transitory regime, from relativistic to non-relativistic, for a DFG.

To go further we must now expand $f$ about its zeroth-order value as
\begin{eqnarray}\label{f_dfg}
f\bb{p\bb{E, n_{\jj}},  x^{\ii}, \tau}  &\approx& g \, \left(\exp{\left[\frac{E}{T(1 + \Delta\bb{x^{\ii}, n_{\jj}, \tau})} - \frac{\mu_{\0}}{T_{\0}}\right]} + 1\right)^{\mi\1} \nonumber\\
&\approx& f\bb{E, \tau} (1 +   \Psi\bb{E, x^{\ii}, n_{\jj}, \tau}),
\end{eqnarray}
where the perturbed piece is concerned with our {\em second approximation} since it can be written as
\begin{equation}
\Psi\bb{E, x^{\ii}, n_{\jj}, \tau}
= \frac{T}{f\bb{E, \tau}} \frac{\partial f}{\partial T} \, \Delta \bb{x^{\ii}, n_{\jj}, \tau}
=- \frac{1}{f\bb{E, \tau}} \frac{E^{\2}}{p} \frac{\partial f}{\partial p} \, \Delta \bb{x^{\ii},n_{\jj}, \tau}.
\label{psipsi}
\end{equation}

The general expression for the energy-momentum tensor written in terms of the distribution function and the momentum components is given by
\begin{equation}
\label{tmunu}
	T^{\mu}_{\nu} = \int dp_{\1} dp_{\2} dp_{\3}\,(-Det[g_{\mu\nu}])^{\mi\1/\2}\,
	{p^\mu p_\nu\over p^{\0}} f\bb{p\bb{\epsilon, n_{\jj}},  x^{\ii}, \tau},
\end{equation}
In the conformal Newtonian gauge, $(-Det[g_{\mu\nu}])^{\mi\1/\2} = a^{\mi\4}(1-\psi+3\phi)$, $dp_{\1} dp_{\2} dp_{\3} = (1 - 3\phi) q^{\2} \,dq \,d\Omega$, and $p_{\0}=-(1 + \psi) \epsilon$.
Keeping $\epsilon$ as $a\bb{\tau}$ times the proper energy measured by a comoving observer and introducing the {\em tilde}-variables as
\begin{equation}
T^{\mu}_{\nu} = \int{{d\Omega \over (2\pi)^{\3}}\tilde{T}^{\mu}_{\nu}},
\end{equation}
to linear order in the perturbations one writes
\begin{eqnarray}
\tilde{T}^{\0}_{\0}   &=-&  \int_{_{0}}^{^{\infty}}{dp\, p^{\2}\, E \,  f\bb{E, \tau} (1 + \Psi)},\nonumber\\
\tilde{T}^{\0}_{\ii}  &=&  \int_{_{0}}^{^{\infty}}{dp\, p^{\3}\,  f\bb{E, \tau} \Psi}\, n_{\jj},\nonumber\\
\tilde{T}^{\ii}_{\jj} &=&  \int_{_{0}}^{^{\infty}}{dp\, \frac{p^{\4}}{E}\,  f\bb{E, \tau} (1 + \Psi)}\,n^{\ii}n_{\jj}.
\label{tmunu2}
\end{eqnarray}
Noticing that the angular dependence in terms of $x^{\ii}$ and $\hat{n}$ can be decoupled from the $q$-integration through Eq.~(\ref{psipsi}), the {\em tilde}-perturbations to the energy-momentum tensor can be written as
\begin{eqnarray}
\delta\tilde{T}^{\0}_{\0}   &=&  \left[\int_{_{0}}^{^{\infty}}{dp\, p     \,E^{\3} \frac{\partial f}{\partial p}} \right]\, \Delta\bb{x^{\ii},n_{\jj}, \tau},          \nonumber\\
\delta\tilde{T}^{\0}_{\ii}  &=-&  \left[\int_{_{0}}^{^{\infty}}{dp\, p^{\2}\,E^{\2} \frac{\partial f}{\partial p}} \right]\, \Delta\bb{x^{\ii},n_{\jj}, \tau}\, n_{\jj},\nonumber\\
\delta\tilde{T}^{\ii}_{\jj} &=-&  \left[\int_{_{0}}^{^{\infty}}{dp\, p^{\3}\,E      \frac{\partial f}{\partial p}} \right]\, \Delta\bb{x^{\ii},n_{\jj}, \tau}\,n^{\ii}n_{\jj}
\label{tmunu3}
\end{eqnarray}

One can notice that we have eliminated the explicit dependence on the metric perturbations.
Rewriting the integrands in terms of the (pseudo)comoving energy, $\epsilon = a\, E = a\, (p^{\2}+m^{\2})^{\1/\2} =  (q^{\2} + a^{\2}m^{\2})^{\1/\2}$, and of the comoving momentum, $q = p\,a$, and integrating by parts, it results in the following expressions for the {\em tilde}-perturbations to the energy-momentum tensor
\begin{eqnarray}
\delta\tilde{T}^{\0}_{\0}   &= -&  a\bb{\tau}^{\mi\4} \left[m^{\2}a\bb{\tau}^{\2} \int_{_{0}}^{^{\infty}}{dq\, \epsilon \,         f\bb{E, \tau}} + 4                      \int_{_{0}}^{^{\infty}}{dq\,q^{\2}\, \epsilon \, f\bb{E, \tau}} \right]\, \Delta\bb{x^{\ii},n_{\jj}, \tau},\nonumber\\
\delta\tilde{T}^{\0}_{\ii}  &=&   a\bb{\tau}^{\mi\4} \left[4                     \int_{_{0}}^{^{\infty}}{dq\, q^{\3} \,           f\bb{E, \tau}} + 2 m^{\2}a\bb{\tau}^{\2}\int_{_{0}}^{^{\infty}}{dq\,q \, f\bb{E, \tau}} \right]\, \Delta\bb{x^{\ii},n_{\jj}, \tau}\, n_{\jj},\nonumber\\
\delta\tilde{T}^{\ii}_{\jj} &=&  a\bb{\tau}^{\mi\4} \left[3                     \int_{_{0}}^{^{\infty}}{dq\,q^{\2}\, \epsilon \, f\bb{E, \tau}} +                        \int_{_{0}}^{^{\infty}}{dq\,\frac{q^{\4}}{\epsilon}\, f\bb{E, \tau}} \right]\, \Delta\bb{x^{\ii},n_{\jj}, \tau}\,n_{\ii}n_{\jj}
\label{tmunu4}
\end{eqnarray}
where it is easy to identify the average energy density and pressure respectively as
\begin{equation}\label{rho_dfg}
\rho\bb{\tau} =  \frac{1}{2 \, \pi^{\2} a\bb{\tau}^{\4}} \int_{_{0}}^{^{\infty}}{dq\,q^{\2}\, \epsilon \, f\bb{E, \tau}},
\end{equation}
and
\begin{equation}\label{p_dfg}
P\bb{\tau} = \frac{1}{2 \,\pi^{\2} a\bb{\tau}^{\4}} \int_{_{0}}^{^{\infty}}{dq\,\frac{q^{\4}}{3\epsilon}\, f\bb{E, \tau}},
\end{equation}
and where $\Delta\bb{x^{\ii},\hat{n}_{\jj}, \tau}$ reads ${\mathcal N}\bb{k^{\ii},\hat{n}_{\jj}, \tau}$ in the Fourier space, i. e.
\begin{equation}
\Delta\bb{x^{\ii},\hat{n}_{\jj},\tau} = \frac{1}{(2\pi)^{\3}}
\int{dk^{\3}\, \exp{[i k_{\ii} x^{\ii}]}\, {\mathcal N}\bb{k^{\ii},\hat{n}_{\jj}, \tau}}.
\end{equation}

When particles are ultra-relativistic, i. e. $\epsilon = q$, the above results reduce to that obtained for massless particles \cite{Ma94,Pas06}.
Differently from that observed when one sets the ultra-relativistic approximation on $f\bb{E, \tau}$, the spectrum does not remain Planckian.
However, under certain particular circumstances, the local dependence on $\epsilon\bb{q}$ and $q$ remains easy to treat analytically.
The $q$-integration can be performed separately and the dependence on $x^{\ii}$, $\hat{n}$ and $\tau$ is kept active.

Now, the dimensionality of the problem can be reduced by noticing that the evolution depends only on the direction $\hat{n}$ through the angle related to $\hat{k} \cdot \hat{n} = \cos\varphi$, which is a natural consequence of the isotropy of the homogeneous background.
We perform the ${\mathcal P}_{\s}$-Legendre expansion with respect to $\cos(\varphi)$,
\begin{equation}
{\mathcal N} \bb{x^{\ii},\hat{n}_{\jj}, \tau} = \sum_{s=0}^{\infty} (-1)^{\s} (2s+1) N_{\s}\bb{k^{\ii}, \tau} {\mathcal P}_{\s}\bb{\cos\varphi},
\end{equation}
that results in
\begin{equation}
N_{\0} = \frac{1}{4\pi} \int{d\Omega\,{\mathcal N}}, ~~ N_{\1} = \frac{i}{4\pi} \int{d\Omega\,\cos\varphi\,{\mathcal N}}~~ \mbox{and}~~N_2=-\frac{3}{8\pi} \int{d\Omega\,\left(\cos^{\2}\varphi - \frac{1}{3}\right)\,{\mathcal N}},
\end{equation}
for the first three $n$-poles.

Introducing the above results after performing the angular integrations over the {\em tilde} variables, the Eqs.~(\ref{tmunu4}) result in
\begin{eqnarray}
\delta T^{\0}_{\0}   &\Rightarrow& \delta \rho =   N_{\0} \left(4\rho\bb{\tau} +  \frac{m^{\2}}{2 \pi^{\2} a^{\2}}\int_{_{0}}^{^{\infty}}{dq\, \epsilon \, f\bb{E, \tau}}\right),\nonumber\\
i\, k^{m}\delta T^{\0}_{m}  &\Rightarrow&  (\rho\bb{\tau} + P\bb{\tau}) \theta =
\frac{k \, N_{\1}}{2\pi^{\2} a^{4}}
\left[4\int_{_{0}}^{^{\infty}}{dq\, q^{\3} \, f\bb{E, \tau}} + 2m^{\2}a^{\2}\int_{_{0}}^{^{\infty}}{dq\,q \, f\bb{E, \tau}} \right],\nonumber\\
-(\hat{k}_{\ii}\hat{k}^{\jj} - (1/3)\delta_{\ii}^{\jj})\Sigma^{\ii}_{\jj} &\Rightarrow& (\rho\bb{\tau} + P\bb{\tau}) \sigma = 2 N_{\2}(\rho\bb{\tau} + P\bb{\tau}).
\label{tmunu5}
\end{eqnarray}

The results of this section allows one to perform preliminary analytical computations for the evolution of perturbations for several classes of relativistic and non-relativistic test fluids.
As an example, in the section IV we shall obtain such results for a DFG in the RD universe and some extensions to MD and $\Lambda$D scenarios.


\section{Perturbations and Einstein equations for a DFG}

The {\em third approximation} is the simplest one.
It sets the DFG cut-off on the distribution function $f\bb{E,\, \tau}$, which results in obtaining analytical expressions for the elements described in Eq.~(\ref{tmunu4}).
In the literature \cite{Pas06,Ma94,Dodelson}, however, one can notice that $f\bb{E,\, \tau}$ is often approximated by $f\bb{p} = (\exp{[p/T]} + 1)^{\mi\1}$.
Besides not resulting in the expected smooth transition between relativistic to non-relativistic regimes, it is out of the scope of the study of a DFG cosmological test fluid. From such a point, the perturbations to a DFG can be analytically expressed without additional assumptions over $f\bb{E,\, \tau}$.

In the limit where $T$ tends to $0$ in Eq.~(\ref{f_dfg}), the Fermi distribution $f\bb{E,\, \tau}$ becomes a step function that yields an elementary integral with the upper limit equal to the (comoving) Fermi momentum, $q_{F}$.
Introducing the variable $y\bb{a} = \lambda / a$, with $\lambda = q_{\F}/ m = (q_{\F}/T_{\0})/(m/T_{\0})$ to parameterize the kinetic energy with respect to the rest mass, the equation of state of the DFG can be expressed in terms of elementary functions of $y\equiv y\bb{a\bb{\tau}}$ as
\begin{eqnarray}
\rho \bb{y} &=& \frac{g\;m^{\4}}{16\;\pi^{\2}}
\left[y (2 y^{\2} + 1)\sqrt{y^{\2} + 1} -
\mbox{arc}\sinh\bb{y}\right],\nonumber\\
P\bb{y} &=& \frac{g\;m^{\4}}{16\;\pi^{\2}}
\left[y (\frac{2}{3} y^{\2} - 1)\sqrt{y^{\2} + 1} +
\mbox{arc}\sinh\bb{y}\right],\nonumber\\
n\bb{y} &=& \frac{g\;m^{\3}}{6\; \pi^{\2}} y^{\3}.
\label{star05}
\end{eqnarray}
from which ultra-relativistic and non-relativistic regimes can thus be analytically connected during the cosmological evolution.

The pressure and energy density are then related by
\begin{equation}
\frac{\partial\rho\bb{y}}{\partial (m\, n\bb{y})}  = \frac{\rho\bb{y} + P\bb{y}}{m\,n\bb{y}} = \sqrt{y^{\2} + 1}.
\label{star06}
\end{equation}

From Eqs.~(\ref{star05}), the relations of the density fluctuation, $\delta$, the fluid velocity divergence, $\theta$, and the shear stress, $\sigma$, with the multipoles of lower order, $N_{\0}$, $N_{\1}$ and $N_{\2}$ can be then be rewritten as
\begin{eqnarray}
\delta \rho &=& N_{\0}y(1+y^{\2})^{\3/\2}\frac{m^{\4}}{\pi^{\2}}, \nonumber\\
\theta &=& 3 k\, N_{\1}\frac{\sqrt{1+y^{\2}}}{y}, \nonumber\\
\sigma &=& 2  N_{\2},
\label{star07}
\end{eqnarray}
as well as
\begin{equation}
\delta P  =  N_{\0}y^3\sqrt{1+y^{\2}}\frac{m^{\4}}{3\pi^{\2}}.
\end{equation}

\begin{figure}
\vspace{-2.5 cm}\centerline{\psfig{file= 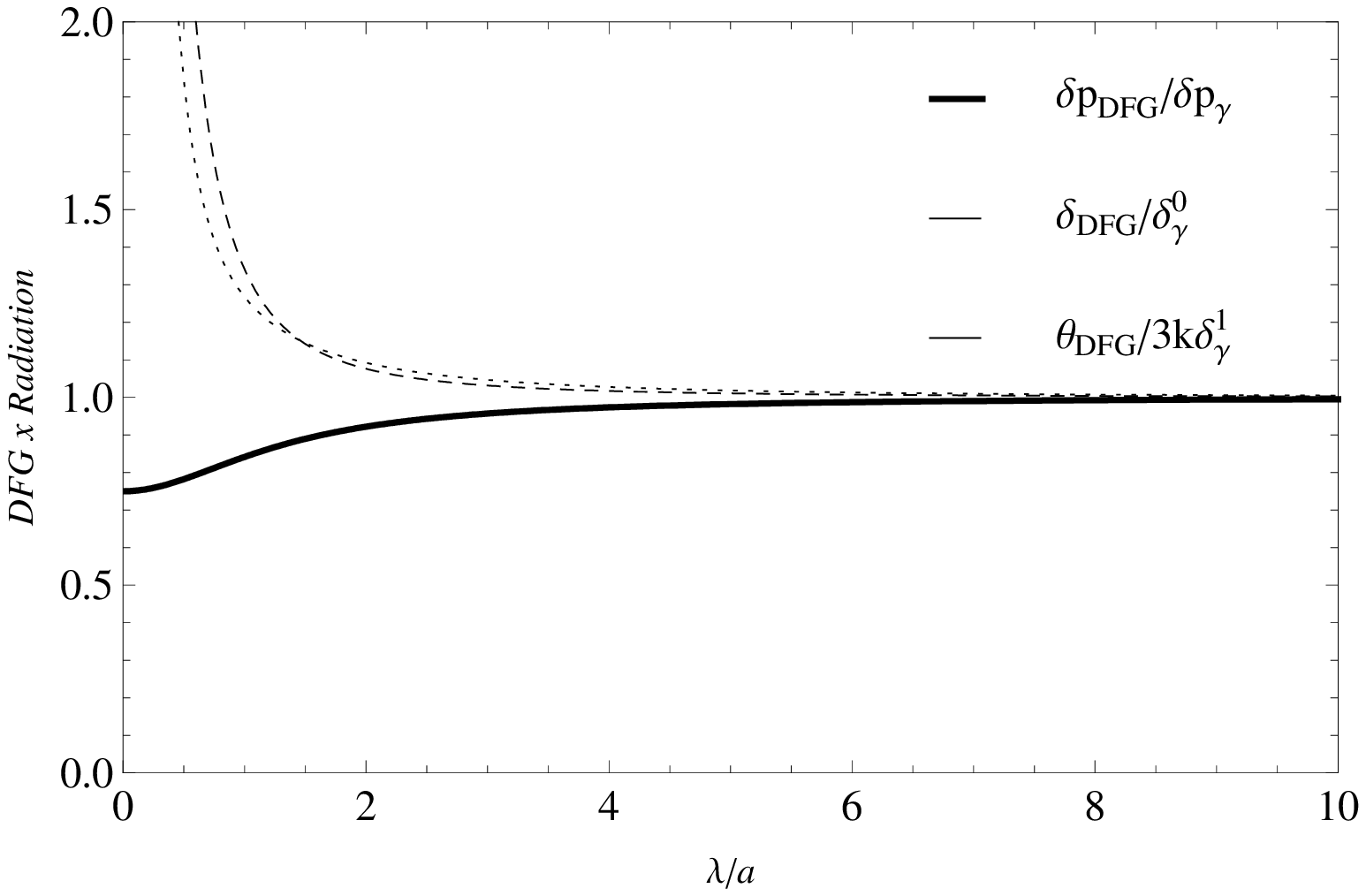,width=13 cm}}
\vspace{-2.5 cm}\centerline{\psfig{file= 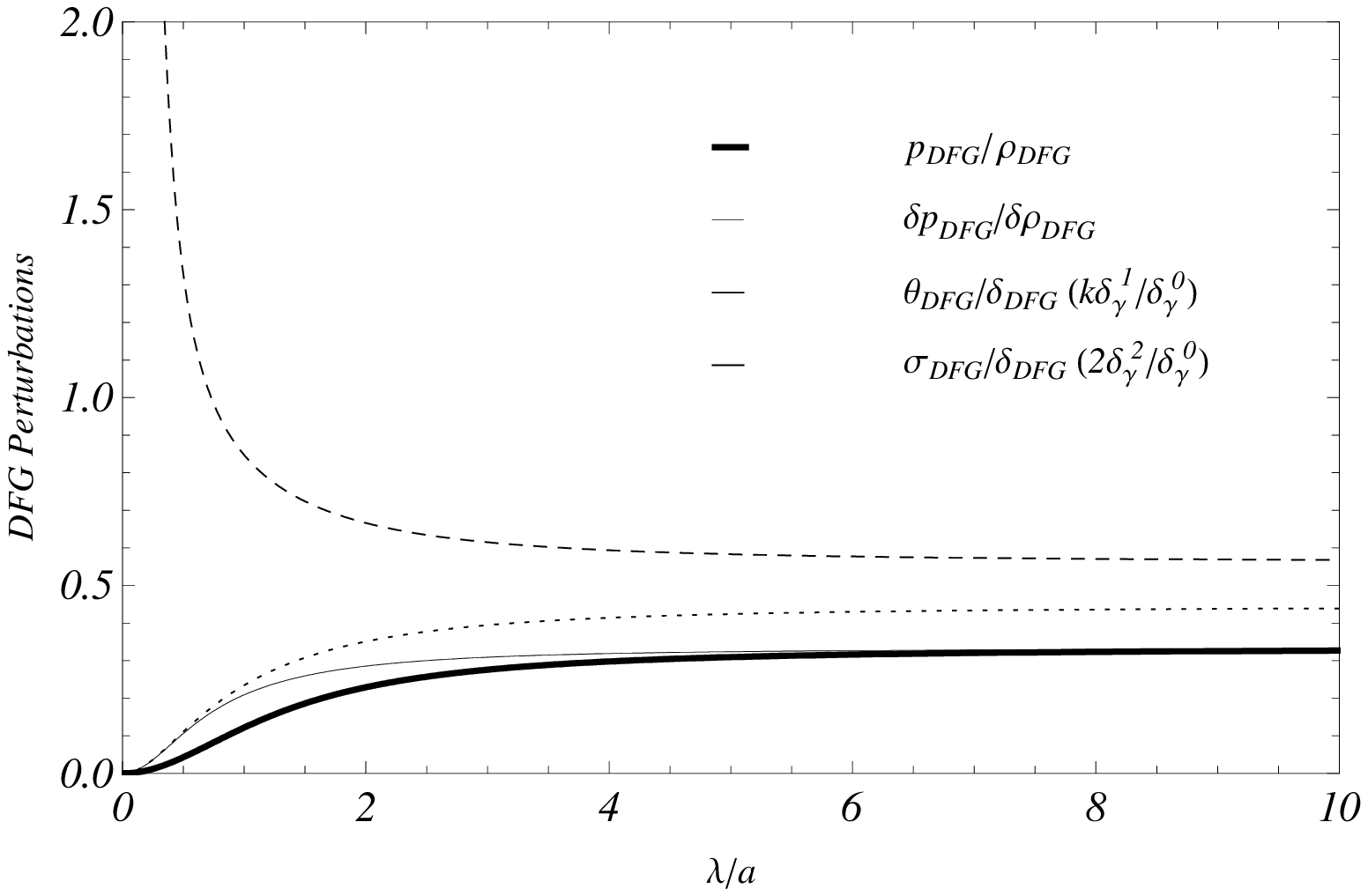,width=13 cm}}
\vspace{-3.0 cm}
\caption{Evolution of the DFG relative isotropic pressure perturbation $\delta P/\rho$, fluid velocity divergence, $\theta$, and anisotropic stress, $\sigma$, in units of the density perturbation, $\delta$, in dependence on the inverse of the scale factor, $a^{\mi\1}$, weighed by the DFG factor $\lambda$.
In the non-relativistic limit, i. e. when $\lambda \ll 1$, one recovers the values for matter perturbation.
In the ultra-relativistic limit, i. e. when $\lambda \gg 1$, one recovers the values for radiation perturbation.
The comparison to radiation perturbations follows in the second plot.}
\label{Figura1}
\end{figure}

In Fig.\ref{Figura1} we describe the most relevant perturbation variables for a DFG, by quantifying their relative magnitudes with respect to the energy density fluctuation, and by comparing it with the results for radiation perturbation.

By introducing the above simplifications given by Eqs.~(\ref{star05}) into Eqs.~(\ref{delta2},\ref{theta2}) for the evolution of perturbations one thus obtains
\begin{equation}
 \frac{d}{d\tau}\theta+\theta\left[aH\left(1-3\frac{P\bb{y\bb{\tau}}}{\rho\bb{y\bb{\tau}}}\right)+\frac{d}{d\tau}\ln\left(1+\frac{P\bb{y\bb{\tau}}}{\rho\bb{y\bb{\tau}}}\right)\right]=\frac{P\bb{y\bb{\tau}}}{\rho\bb{y\bb{\tau}}+P\bb{y\bb{\tau}}}k^{\2} \delta+k^{\2} \phi \label{star08},
\end{equation}
and
\begin{equation}
\frac{\rho\bb{y\bb{\tau}}}{\rho\bb{y\bb{\tau}} + P\bb{y\bb{\tau}}} \frac{d}{d\tau}\delta + \theta= 3\frac{d}{d\tau}\phi\bb{\tau}
\label{star09},
\end{equation}
through which, under simplified assumptions for the background cosmological scenarios that set the dynamical behavior of the Hubble parameter, $H\bb{a}$, we obtain the DFG test fluid evolution.

One should obtain the same results departing from the Boltzmann equation in $k$-space,
\medskip
\begin{equation}
\label{bolt-con}
  {\partial \Psi \over \partial \tau} + i\,{q \over
	\epsilon}\,(\vec{k}\cdot \hat{n})\,\Psi +
     {d\ln f_{\0} \over d\ln q} \left[\dot{\phi} - i\,{\epsilon\over q}
       (\vec{k}\cdot\hat{n})\,\psi \right] =
   {1\over f_{\0}}\,\left( {\partial f \over \partial\tau} \right)_C \,
\end{equation}
once the condition given by (\ref{cond}) had been set.
For instance, the equation for the shear stress, $\sigma = 2 N_{\2}$ stays as
\begin{equation}
\frac{\partial}{\partial\tau}\left(\frac{\epsilon^{\2}}{q}N_{\1}\right) + k \frac{\epsilon}{5}\left[\sigma - N_{\0}\right] = 0
\label{star10}.
\end{equation}

\section{DFG test fluid perturbations}

Deep inside the RD era, one sees that the radiation component can be approximated by a perfect fluid of sound speed $c_s$, due to the tight coupling between baryon and photons \cite{Dodelson}.
This self-gravitating fluid oscillates inside the sound horizon. The cold dark matter evolves as a test fluid
in this background, until the time (close to equality) at which its gravitational back-reaction becomes important.
In this limit, and in the absence of anisotropic stress ($\phi \equiv \psi$), it is possible to derive an analytic expression for the evolution of the metric perturbation $\phi_\gamma$.
The full continuity, Euler and Einstein equations for the  RD background results in the following expression for the scalar perturbation (see, for instance, \cite{Pas06,Ma94})
\begin{equation}
\phi_\gamma\bb{\omega\tau} = \frac{1}{(\omega\tau)^2}\left[C_\gamma\bb{\omega} \left(\frac{\sin{[\omega\tau]}}{\omega\tau} -\cos{[\omega\tau]}\right) - D_\gamma\bb{\omega}\left( \frac{\cos{[\omega\tau]}}{\omega\tau} +\sin{[\omega\tau]}\right)
\right],\end{equation}
where $\omega$ is proportional to the Fourier wavenumber $k$, $\omega = k/\sqrt{3}$, and $\tau$ is the conformal time related with the proper time through the scale factor by $d\tau = dt/a\bb{t}$ \cite{Pas06}.
We just have noticed that the simplest way to describe the subsequent results is in terms of the dimensionless parameter $\omega\tau$, i. e. the conformal time weighed by $\omega$.

The solution proportional to $C$ represents the growing adiabatic mode, the other one  proportional to $D$ is the decaying adiabatic mode, which is suppressed during the cosmological evolution.

\begin{figure}
\centerline{\psfig{file= 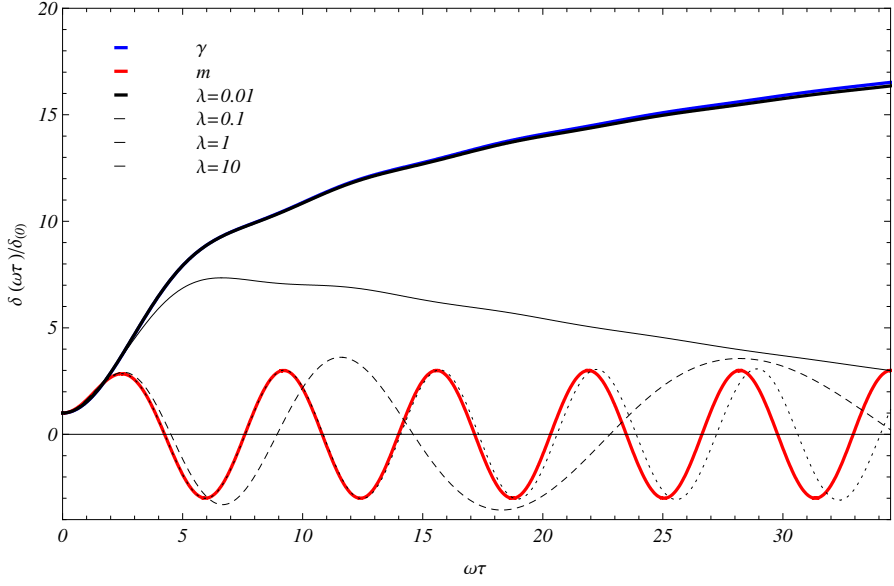,width=13 cm}}
\caption{Evolution of radiation ($\delta_\gamma$), matter ($\delta_m$) and DFG ($\delta_{\lambda}$) energy density fluctuations in dependence on the weighed conformal time, $\omega\tau$, during the RD regime.
The parameter $\lambda$ summarizes the dependence on the relativistic regime and on the cosmological scale.
By varying $\lambda$ one describes how relativistic is the regime; and by varying $\omega$ one sets the relation to the cosmological scale.
The results are for $\lambda = 0.01,\, 0.1,\, 1$ and $10$.
The initial condition was set arbitrary since we are depicting only relative values.}
\label{Figura2}
\end{figure}

\begin{figure}
\centerline{\psfig{file= 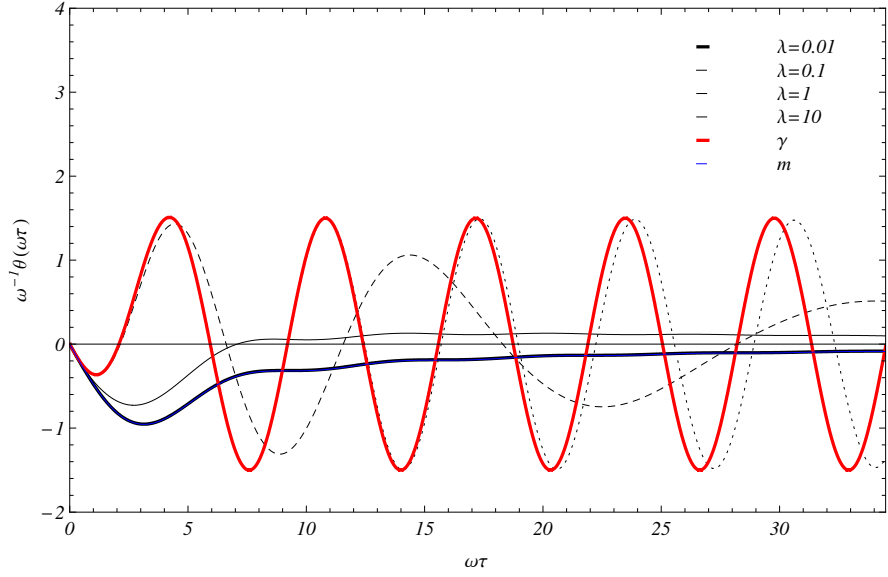,width=13 cm}}
\caption{The fluid velocity divergence, $\theta$, in dependence on the weighed conformal time, $\omega\tau$, during the RD regime.
The results are also for $\lambda = 0.01,\, 0.1,\, 1$ and $10$.}
\label{Figura3}
\end{figure}

By substituting the above expression for $\phi_\gamma$ into the Eqs.~(\ref{star08})-(\ref{star09}) one can obtain $\delta_{\lambda}$ and $\theta_{\lambda}$ numerically for a DFG test fluid.
For preliminary results, we have approximated the Universe composition by $\Omega_\gamma = 0.01$, $\Omega_m = 0.29$, and $\Omega_\Lambda = 0.70$.
In order to perform a generical analysis in terms of non-relativistic to ultra-relativistic regimes ($\lambda$) and cosmological scales ($\omega\tau$), we have considered different values for the parameter $\lambda$ in resolving Eqs.~(\ref{star08})-(\ref{star09}).
The results can be depicted from Figs.~\ref{Figura2} and \ref{Figura3} where the dependence on the relativistic regime through $\lambda$ does not obey a regular behavior, i. e. the most subtle effect on the DFG cosmic fluid is the suppression of the growing modes while the RD universe is cooling.
In some sense it is unnatural since one should expect DFG approximating the non-relativistic matter behavior (growing modes) as the universe cools.
We compare the results to those for matter (blue line) and radiation (red line) obtained from direct resolution of Eqs.~(\ref{delta2}-\ref{theta2}), respectively in the NR and UR limits,
\begin{eqnarray}
	\dot{\delta_m} &=& - \left(\theta_m - 3 {\dot{\phi}}\right),\nonumber\\
	\dot{\theta_m} &=& - {\dot{a}\over a} \theta_m + 3 \omega^{\2} \phi,
\label{matter}
\end{eqnarray}
and
\begin{eqnarray}
	\dot{\delta_{\gamma}} &=& - \frac{4}{3}\left(\theta_{\gamma} + 4 {\dot{\phi}}\right),\nonumber\\
	\dot{\theta_{\gamma}} &=& \frac{3}{4} \omega^{\2} \delta_{\gamma} + 3 \omega^{\2} \phi.
\label{radiation}
\end{eqnarray}
Turning back to the relation between the conformal time, $\tau$, and the scale factor, $a\bb{\tau}$, for the RD era, one has
\begin{equation}
a_{\gamma}\bb{\tau} \sim \Omega_{\gamma}^{\1/\2} \, \tau,
\end{equation}
and, by adopting adequate continuity conditions, one sets
\begin{equation}
a_{m}\bb{\tau} \sim \frac{\Omega_m}{4}\left( \tau + \frac{\Omega_{\gamma}^{\1/\2}}{\Omega_m}\right)^{\2},
\end{equation}
and
\begin{equation}
a_{\Lambda}\bb{\tau} \sim \frac{1}{\Omega_\Lambda^{\1/\2}}
\left(\frac{3}{\Omega_m^{\1/\3}\,\Omega_{\Lambda}^{\1/\6}} - \frac{\Omega_{\gamma}^{\1/\2}}{\Omega_m}-\tau\right)^{\mi\1},
\end{equation}
respectively for the MD and $\Lambda$D eras.
One thus finds the conformal time for the equality between radiation and matter densities, $\tau_{eq}$, for the equality between matter and cosmological constant $\Lambda$ densities, $\tau_{m\Lambda}$, and for the present $\Lambda$ dominance at $a = 1$, $\tau_{\Lambda 0}$,
\begin{eqnarray}
\tau_{eq}        &\sim& \frac{\Omega_{\gamma}^{\1/\2}}{\Omega_m}, \nonumber\\
\tau_{m\Lambda}    &\sim& \frac{2}{\Omega_m^{\1/\3}\,\Omega_{\Lambda}^{\1/\6}} -\tau_{eq}  , \nonumber\\
\tau_{\Lambda 0} &\sim& \frac{3}{2}\tau_m +\frac{1}{2}\tau_{eq} -\frac{1}{\Omega_\Lambda^{\1/\2}}.
\label{times}
\end{eqnarray}

In addition, the analytical solutions for the metric perturbations $\phi\bb{\tau}$ related to the MD and $\Lambda$D regimes, are respectively given by
\begin{eqnarray}
\phi_m\bb{\omega\tau} &=& C_m \bb{\omega} + {D_m\bb{\omega} \over (\omega\tau)^{\5}},
\end{eqnarray}
and
\begin{eqnarray}
\phi_m\bb{\omega\tau} &=& C_\Lambda \bb{\omega} (a_{\Lambda}\bb{\tau})^{\mi\3} + D_\Lambda \bb{\omega} (a_{\Lambda}\bb{\tau})^{\mi\1}.
\end{eqnarray}
By fine-tuning the transition between RD, MD and $\Lambda$D solutions through adequate continuity conditions between the growing adiabatic mode of $\phi_{\gamma}$ and the complete solution for $\phi_{m}$ and between their first derivatives, both at $\tau_{eq}$, and subsequently doing the same for the complete solutions of $\phi_m$ and $\phi_\Lambda$ at $\tau_{m}$, one can easily find the coefficients $C_{m}$, $C_{\Lambda}$, $D_{m}$, and $D_{\Lambda}$.

\begin{figure}
\vspace{-0.5 cm}\centerline{\psfig{file= 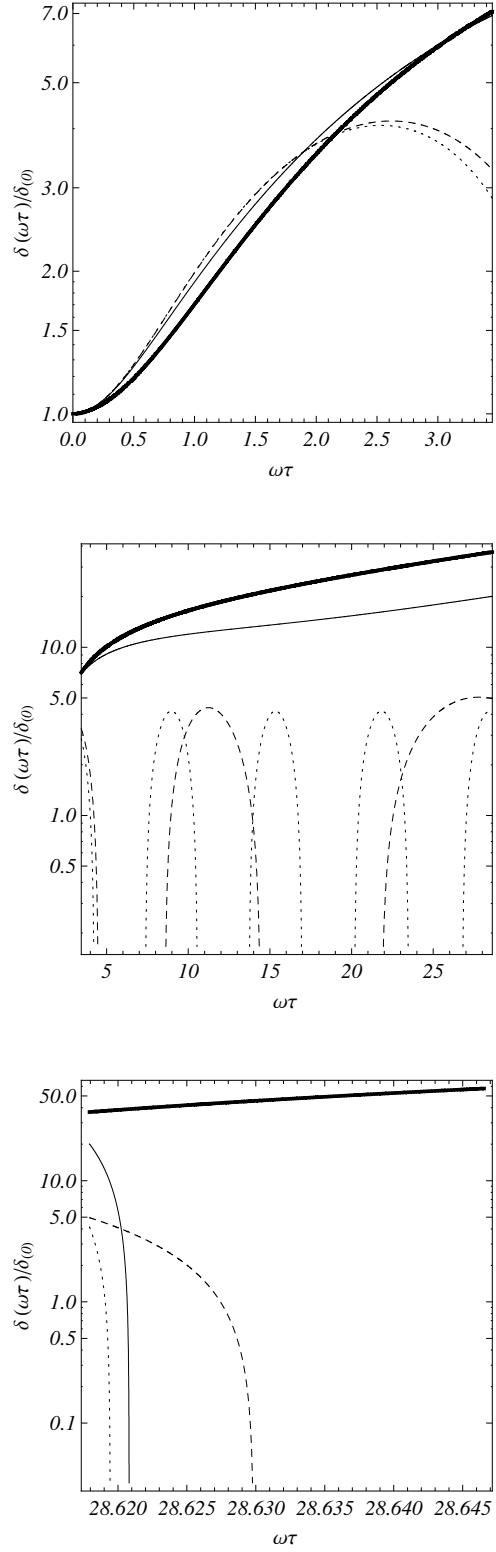,width= 7 cm}}
\caption{The density energy fluctuations, $\delta_{\lambda}$, in dependence on the weighed conformal time, $\omega\tau$, respectively for RD, MD and $\Lambda$D regimes.
Time is evolving from the beginning of the RD era.
The values for the parameter $\lambda$ are in correspondence with the previous figures.}
\label{Figura4}
\end{figure}

This corresponds to the simplest strategy for obtaining the first approximation for a {\em transference function} between radiation, matter and $\Lambda$ eras, since it allows one to build continuous solutions for the density fluctuation $\delta_{\lambda}$ and fluid velocity divergence $\theta_{\lambda}$ of the DFG test fluid.
In the Fig.~\ref{Figura4} we have illustrated the results for $\delta_{\lambda}$, in correspondence with the $\lambda$ values introduced in the previous results for classifying the DFG test fluid according to its relativistic regime.
The cut-off in each plot is in agreement with the values considered in (\ref{times}).

One has to notice that the relation between $\theta$ and $\omega$ are relevant for determining the behavior of the density fluctuations.
The first stage (i. e. at RD era) of growing perturbations depends on how smaller is $\lambda$, i. e. they are favored by non-relativistic configuration regimes of the DFG test fluid.
Otherwise it also depends on scales parameterized by the weight factor $\omega$ in the $x$-axis for the conformal time.
When the perturbation scale is well inside the Hubble horizon, the oscillating behavior (RD) followed by a strong suppression at the last stage ($\Lambda$D) is favored. For large scales, the growing perturbations may or may not be suppressed at the last stage.
Such a qualitative preliminary analysis allows for reheating the discussion on the possibility of forming DFG structures at different cosmological eras, and deserve some attention, since the behavior of DFG perturbations does not mimics matter perturbations exactly, with the exception of the totally NR limit.

\section{Conclusions}

An analytical procedure for obtaining the evolution of perturbations for a class of relativistic and non-relativistic test fluids was discussed.
The density fluctuation, $\delta$, the fluid velocity divergence, $\theta$, and an explicit expression for the dynamics of the shear stress, $\sigma$, were obtained for a DFG in the background regime of radiation, and some complementary results were obtained in the case matter domination and $\Lambda$CDM cosmological eras.
The comparisons with previously known results from radiation and matter perturbations are immediate from our analysis and show the consistence with the DFG approach in the sense that an equivalent equation of state can smoothly transit from the UR to the NR limits.
In particular, one can observe that a DFG, even in the non-relativistic regime at early times preconize the growing of perturbations in the same intensity as matter.
However, for intermediate regimes, the growing adiabatic modes are relatively suppressed after the beginning, turning to an oscillatory behaviour.
Also for the fluid velocity divergence, the transition from NR to UR regimes can be continuously observed.
In this sense, our results become a convenient tool for performing preliminary tests for HDM to CDM test fluid configurations.
Moreover, the DFG approach can be used to set a correspondence between the averaged neutrino mass at present ($a = 1$), thorough the NR limit of the energy density, and the Fermi momentum, $p_F$, through the UR neutrino energy density at the RD era.

Concerning the neutrino degeneracy, one can ascertain that the value of the chemical potential of electron neutrinos ($\mu_e/T_0$) should be close to $-1$ in  order to explain with an acceptable confidence level the phenomenology related to the observed variation of deuterium by roughly an order of magnitude \cite{Dolgov02}.
It would induce a variation in total energy density during the RD stage, which is excluded by the smoothness of CMB.
However, this objection could be avoided if there was a coincidence between different leptonic chemical potentials
such that in different spatial regions they had the same values but with interchange of electronic,
muonic and/or tauonic chemical potentials \cite{Hu93A}.
For further details and discussion of energy dependence and effective chemical potential one might address the papers \cite{Hu93A,Hu93B} where the main point is that they are not obtrusive to our results.

It should also be mentioned that cosmological models with a large excess of neutrinos - or antineutrinos - for example, have been discussed \cite{Bor03} in this context.
In such models there is a large neutrino background behaving like a DFG.
The equilibrium distributions show that for $\mu_\nu \neq 0$, and $|\mu_\nu| \gg k T$, there is an excess of particles.
During the RD phase of the early universe, a non-zero chemical potential for neutrinos, $\mu_\nu$, can give rise to an excess of particles over antiparticles or vice-versa.
The massless neutrinos (or antineutrinos) would then form a Fermi sea with degenerate particles filling the Fermi levels up to $p_F$.
However, the exact neutrino number density is not known as it depends on the unknown chemical potentials for the three flavors. (In addition there could be a population of sterile neutrinos \cite{Aba90}, hypothesis that we did not discuss here.)
In this context it is also frequent to notice the term neutrino stars as a generic name for any degenerate fermion star composed of neutral weakly interacting fermions, e.g., neutrinos or supersymmetric fermionic partners \cite{Vio93,Vio98,Bil01}.

Finally, the hypothesis of a small fraction of dark matter components evolving cosmologically as a DFG test fluid in some dominant cosmological background can also be speculated.
The idea is based on the fact that for any type of fermionic dark matter, the average phase-space density cannot exceed the phase-space density of the DFG.
In this case, instead of specifically considering structures like exotic stars or exotic planets, one should consider the possibility of forming extended overdense regions.
This study can be relevant for massive neutrinos in dark matter models and for collisionless self-gravitating systems experiencing some kind of relaxation.
Our results could thus be extended to support the discussion of the formation of the halo of our galaxy assuming that it consists of a gas of massive degenerate fermions in hydrostatic and thermal equilibrium at finite temperatures.
Such a formulation is consistent with the precursive idea of a self-gravitating Fermi gas model introduced to explain the puzzling nature of white dwarf stars \cite{Fowler1926,Chandra1939}.
This fermion gas model is expected to imply the existence of supermassive compact dark objects at the galactic center.

\begin{acknowledgments}
A. E. B. would like to thank for the financial support from the Brazilian Agencies FAPESP (grant 08/50671-0) and CNPq (grant 300627/2007-6), and for the hospitality of the IFGW - UNICAMP.
\end{acknowledgments}

\pagebreak
\newpage

\end{document}